\documentclass{article}
\usepackage{spconf,amsmath,graphicx}

\usepackage{enumitem}
\usepackage{bm}
\usepackage{amssymb}
\usepackage{multirow}
\usepackage{pifont}
\usepackage{hyperref}

\setlist{nosep, leftmargin=14pt}

\usepackage{mwe} 

\usepackage[]{acronym}
\acrodef{DL}{Deep Learning}
\acrodef{US}{Ultrasound}
\acrodef{PCCL}{{\bf P}ixel-level and {\bf C}lass-level {\bf C}onsistency {\bf L}earning}
\acrodef{CNN}{Convolutional Neural Network}
\acrodef{ROI}{Region of Interest}
\acrodef{SSL}{Semi-Supervised Learning}
\acrodef{GT}{Ground Truth}
\acrodef{CR}{Consistency Regularization}
\acrodef{EMA}{Exponential Moving Average}
\acrodef{MAC}{Mutual Agreement Consistency}
\acrodef{KL}{Kullback–Leibler divergence}
\acrodef{MIG}{Mutual Information Gap}
\acrodef{CE}{Cross-Entropy}
\acrodef{Dice}{Dice based coefficient}

\newcommand{\Dice}{\mathcal{D}}
\newcommand{\CE}{\mathcal{H}}

\newcommand{\MSE}{\mathcal{M}}
\newcommand{\Downsampling}{\mathcal{D}}
\title{Entropy-Guided Agreement-Diversity: A Semi-Supervised Active Learning Framework for Fetal Head Segmentation in Ultrasound}
\name{Author(s) Name(s)\thanks{A footnote can be used to report funding sources and other potential conflicts of interest. Use Times 9-point type, single-spaced.}}
\address{Author Affiliation(s)}
\name{Fangyijie Wang$^{\star \ddagger \circ}$\thanks{$\star$ Corresponding author (fangyijie.wang@ucdconnect.ie)} \thanks{This work was funded by Taighde \'{E}ireann – Research Ireland through the Centre for Research Training in Machine Learning (18/CRT/6183).} \qquad Siteng Ma$^{\dagger}$ \qquad Gu\'enol\'e Silvestre$^{\ddagger \dagger}$ \qquad Kathleen M. Curran$^{\ddagger \circ}$}
\address{$^{\ddagger}$ Taighde \'{E}ireann – Research Ireland Centre for Research Training in Machine Learning, Ireland \\
$^{\dagger}$ School of Computer Science, University College Dublin, Ireland \\
$^{\circ}$ School of Medicine, University College Dublin, Ireland \\
}

\begin{document}
%
\maketitle

\begin{abstract}
Fetal ultrasound (US) data is often limited due to privacy and regulatory restrictions, posing challenges for training deep learning (DL) models. While semi-supervised learning (SSL) is commonly used for fetal US image analysis, existing SSL methods typically rely on random limited selection, which can lead to suboptimal model performance by overfitting to homogeneous labeled data. To address this, we propose a two-stage Active Learning (AL) sampler, \textbf{E}ntropy-\textbf{G}uided \textbf{A}greement-\textbf{D}iversity (EGAD), for fetal head segmentation.
Our method first selects the most uncertain samples using predictive entropy, and then refines the final selection using the agreement-diversity score combining cosine similarity and mutual information.
Additionally, our SSL framework employs a consistency learning strategy with feature downsampling to further enhance segmentation performance. 
In experiments, SSL-EGAD achieves an average Dice score of 94.57\% and 96.32\% on two public datasets for fetal head segmentation, using 5\% and 10\% labeled data for training, respectively. Our method outperforms current SSL models and showcases consistent robustness across diverse pregnancy stage data. The code is available on \href{https://github.com/13204942/Semi-supervised-EGAD}{GitHub}.

\end{abstract}

\vspace*{-0.6em}
\section{Introduction}
Ultrasound (US) imaging is extensively utilized for prenatal assessment of fetal growth, fetal anatomy, estimating gestational age, and monitoring pregnancy due to its portability, affordability, and non-invasive nature~\cite{Salomon:2011,Zaffino:2020}. 
In particular, the analysis of anatomical structures is a critical component of fetal screening as it provides direct evidence of potential fetal malformations, placental localization disorders, and the risk of premature birth~\cite{Salomon:2011}.

Deep learning (DL) has shown great potential for fetal US image analysis~\cite{Plotka2021,Yasrab:2024}. However, robust models require extensive expert annotations, which are costly and time-consuming~\cite{Piaggio:2021,Zegarra:2023}.
A sufficient amount of expert annotations is crucial for developing robust DL algorithms. However, annotating US images is a labor-intensive and time-consuming process that demands medical proficiency and clinical expertise for accurate pixel-level labeling~\cite{Piaggio:2021,Zegarra:2023}. Recent works~\cite{Jiang:2024,Ma:2024,Wang:2025b} have applied semi-supervised learning (SSL) to this field, yet existing methods still struggle with accurate fetal anatomical segmentation, highlighting the need for tailored SSL approaches.


Moreover, current SSL methods typically rely on fixed, randomly selected labeled samples. This often leads to insufficient model learning and introduces prediction-level noise, resulting in suboptimal segmentation performance for small fetal anatomical structures in early-stage ultrasound scans. In contrast, Active Learning (AL) is a machine learning paradigm that can iteratively query the most informative samples for labeling, thereby achieving performance comparable to fully supervised learning with minimal labeled data \cite{Wang:2017}, and addressing the limitations above.

In this paper, we investigate the following research question: {\it Can the AL query strategy help SSL framework improve the segmentation performance of lightweight DL in US analysis?} To answer this question, we propose a novel approach named \textbf{E}ntropy-\textbf{G}uided \textbf{A}greement-\textbf{D}iversity strategy for \textbf{S}emi-\textbf{S}upervised \textbf{L}earning (SSL-EGAD) 
for fetal head segmentation in US. Unlike existing SSL methods, SSL-EGAD is the first algorithm to adapt AL query strategy based on the learning outcomes of the semi-supervised lightweight DL model. 
Our experiments demonstrate that SSL-EGAD outperforms the state-of-the-art (SOTA) methods in fetal head segmentation. This advancement facilitates a cost-effective reduction in labeling costs and enhances the robustness of models for various fetal anatomical segmentation scenarios.

\vspace*{-0.7em}
\section{Methodology}
\begin{figure}
\begin{center}
\includegraphics[width=.9\linewidth]{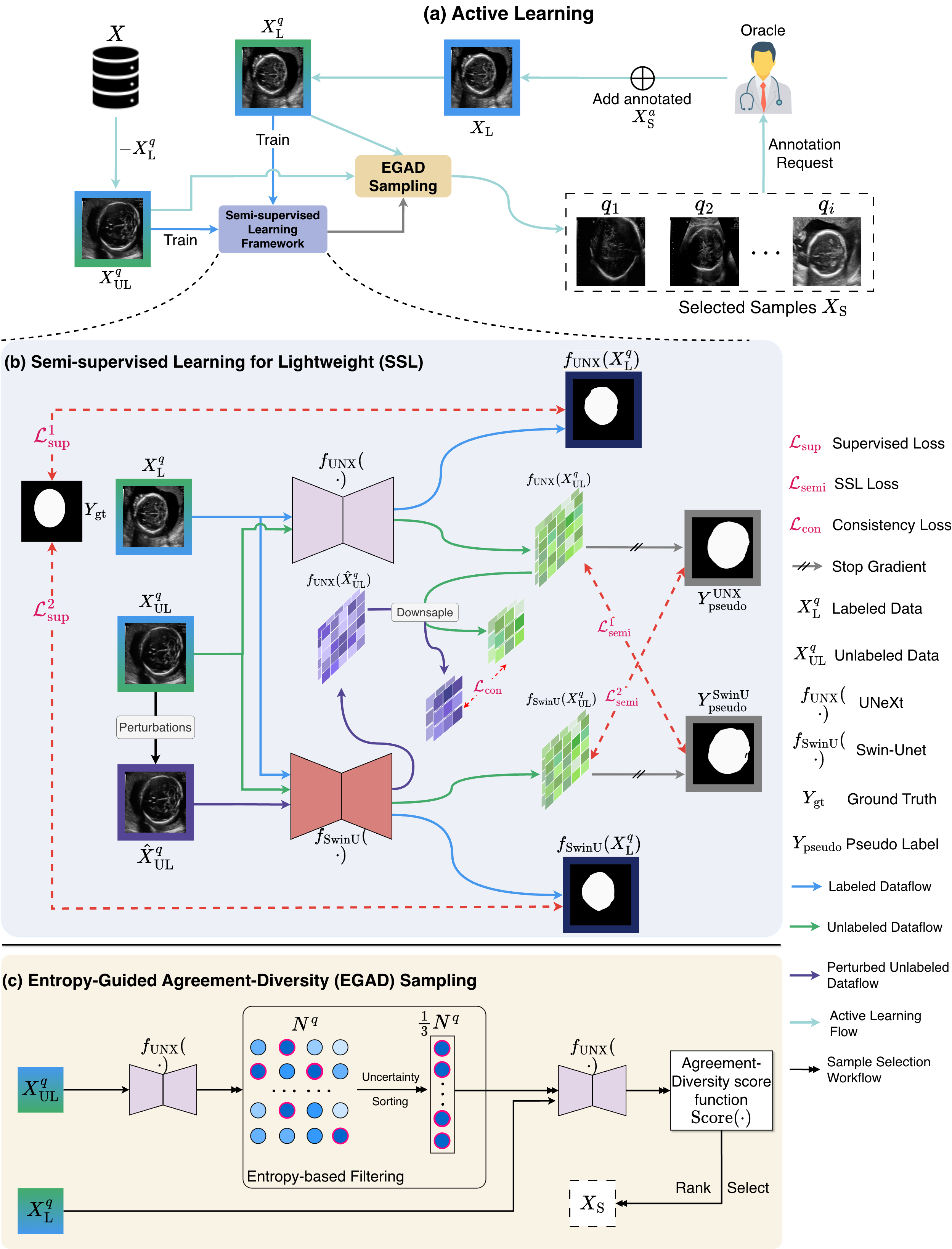}
\end{center}
\vspace{-.7em}
   \caption{An overview of our proposed SSL-EGAD with Swin-Unet and UNeXt for fetal US image segmentation.}
\label{fig:framework}
\vspace{-1em}
\end{figure}

The overview of our method SSL-EGAD is shown in Fig.~\ref{fig:framework}. 
Given a fetal US dataset, 
($\bm X_\text{L} \in \mathbb{R}^{3 \times n_h \times n_w}$, $\bm Y_\text{gt} \in \mathbb{R}^{2 \times n_h \times n_w}$) denotes the initialized labeled training dataset that is randomly selected, while $\bm X_\text{UL}$ denotes the initialized unlabeled training dataset. $\bm X_\text{L}, \bm X_\text{UL} \in \mathbb{R}^{3 \times n_h \times n_w}$ represents a 2D fetal US image of size $n_h\times n_w$ with 3 channels. $\bm X^i_\text{S}$ are selected samples with query strategy $Q$ at each AL round $i$, see details in Section~\ref{al}. We incorporate $\bm X_\text{S}$ into the labeled set $\bm X_\text{L}$ and exclude $\bm X_\text{S}$ from the unlabeled set $\bm X_\text{UL}$ to derive the sets $\bm X^q_\text{L}$ and $\bm X^q_\text{UL}$, respectively. The sets $\bm X^q_\text{L}$ and $\bm X^q_\text{UL}$ are training data of SSL framework.

\vspace{-.7em} 
\subsection{Entropy-Guided Agreement-Diversity Sampler}
\label{al}

To efficiently identify the most informative unlabeled images for annotation, we propose an Entropy-Guided Agreement-Diversity (EGAD) Sampler, illustrated in Fig.~\ref{fig:framework}(c). EGAD adopts a two-stage filter–refine strategy that balances model uncertainty and sample diversity, enabling label-efficient SSL. In the first stage, we measure the predictive uncertainty of each unlabeled image  using the mean predictive entropy of the segmentation probabilities: 
\vspace*{-0.8em}
\[
\mathcal{H}_j = -\sum^{2}_{c=1} \sum_{h=1}^{n_h} \sum_{w=1}^{n_w} {\bm p}_{j,c,h,w} \log \left( {\bm p}_{j,c,h,w} \right)
\]
where ${\bm p}_j \in \mathbb{R}^{2 \times n_h \times n_w}$ is the model prediction. The top one-third of unlabeled samples with the highest entropy values are retained as uncertain candidates, ensuring that subsequent selection focuses on regions of the data space where the model exhibits high epistemic uncertainty. In the second stage, EGAD evaluates each image among the uncertain candidates using a hybrid agreement-diversity score, which captures both semantic similarity in feature space and statistical dependence in image space. For an unlabeled image $\bm X^q_\text{UL}$ and a labeled image $\bm X^q_\text{L}$, we compute:
\vspace*{-0.6em}
\[
\operatorname{CosSim}(\bm X^q_\text{UL}, \bm X^q_\text{L})=\frac{z(\bm X^q_\text{L})^{\top} z(\bm X^q_\text{UL})}{\|z(\bm X^q_\text{L})\|\|z(\bm X^q_\text{UL})\|}
\]
where $z(\cdot)$ denotes the normalized feature embedding, and the empirical mutual information between the two image intensities is defined as:
\vspace*{-0.6em}
\[
\operatorname{MI}\left(\bm X_{U L}^q, \bm X_L^q\right)=\sum_{i, j} p_{\bm X_{U L}^q \bm X_L^q}(i, j) \log \frac{p_{\bm X_{U L}^q \bm X_L^q}(i, j)}{p_{\bm X_{U L}^q}(i) p_{\bm X_L^q}(j)}
\]
The joint distribution $p_{\bm X_{U L}^q \bm X_L^q}$ is estimated via a 2D histogram over the intensity pairs of corresponding pixels across all channels. Each term is normalized to $[0,1]$ across the unlabeled pool, and the final score is:
\vspace*{-0.5em}
\[
\operatorname{Score}\left(X_{U L}^q\right)=\mathcal{N}\left(\operatorname{CosSim}\left(X_{U L}^q\right)\right)-\mathcal{N}\left(\operatorname{MI}\left(X_{U L}^q\right)\right)
\]
The top $k$ samples with the highest scores are selected for annotation. We set $k$ to be one-third. This choice balances uncertainty and efficiency, and EGAD relies on relative ranking rather than the exact threshold, consistent with prior AL practice. 
This two-stage sampling strategy identifies highly uncertain samples and then prioritizes those that are both semantically distinct and statistically independent from the labeled data, yielding a more diverse and informative annotation set. $\operatorname{CosSim(\cdot)}$ and $\operatorname{MI(\cdot)}$ capture complementary redundancy, with normalization ensuring a stable, model-agnostic ranking without explicit weighting.



\vspace*{-0.8em}
\subsection{Semi-supervised Learning via Co-Training}
\label{ssl}
Our proposed SSL method consists of two key components: (1) A co-training strategy for UNeXt~\cite{Valanarasu:2022} and Swin-Unet~\cite{Cao:2023}, represented by $f^{\text{UNX}}_{\theta}(\cdot)$ and $f^{\text{Swin}}_{\phi}(\cdot)$, respectively. (2) A consistency learning method between two networks using pseudo labels, shown in Fig.~\ref{fig:framework}(b).

We utilize UNeXt~\cite{Valanarasu:2022} to construct our lightweight CNN model denoted as $f^{\text{UNX}}_{\theta}(\cdot)$. The number of channels across each block in the encoder path is set to [$32,64,128,160,256$]. 
The $f^{\text{Swin}}_{\phi}(\cdot)$ is implemented using the tiny Swin-Unet architecture from work~\cite{Cao:2023}. To fit the resolution of our input image, which is $448 \times 448$, the shifted window size is set to 7. The bottleneck feature dimension is 768. 
We initialize our model using pre-trained weights~\cite{Cao:2023}.
Table~\ref{model_profile} compares the profiles of UNeXt and Swin-Unet. The UNeXt shows superior lightweight and efficiency to Swin-Unet.

\begin{table}
\centering
\caption{CNN and Transformer models$'$ profiles.}
\resizebox{\linewidth}{!}{%
\label{model_profile}
\setlength{\tabcolsep}{8pt}
\begin{tabular}{|lccc|}
\hline
Model & Input Size & Param (M) $\downarrow$ & GFLOPs $\downarrow$ \\
\hline
Swin-Unet & $448\times448$ & 27.15 & 71.17  \\
UNeXt & $448\times448$ & 1.47 & 7.03  \\
\hline
\end{tabular}%
}
\vspace*{-0.7em}
\end{table}

Building upon the foundation laid by the previous co-training method Cross Pseudo Supervision~\cite{Chen:2021,Ma:2024},
we introduce a simple yet powerful cross-supervision learning framework that incorporates consistency learning. This approach combines a compact CNN network with a Transformer-based network to boost their performance collaboratively. Given the unlabeled dataset $\bm X^q_\text{UL}$ in AL round, the proposed framework produces two predictions: $f^{\text{UNX}}_{\theta}(\boldsymbol{X}^q_\text{UL})$ and $f^{\text{Swin}}_{\phi}(\boldsymbol{X}^q_\text{UL})$.
The pseudo labels for the CPS strategy are generated by:
\vspace*{-0.5em}
\[
\bm \tilde{Y}^{\text{UNX}}=f_\text{OH}\left(f^{\text{UNX}}_{\theta}\left(\boldsymbol{X}_\text{UL}\right)\right);
\bm \tilde{Y}^{\text{Swin}}=f_\text{OH}\left(f^{\text{Swin}}_{\phi}\left(\boldsymbol{X}_\text{UL}\right)\right).
\]
where $f_\text{OH}$ represents a one-hot encoding function. 

Inspired by existing studies~\cite{Tarvainen:2017,Yu:2019}, we leverage consistency learning via image perturbation in training, shown in Fig.~\ref{fig:framework}(b), which helps the $f^{\text{UNX}}_{\theta}(\cdot)$ learn more robust and discriminative features. Specifically, we add Gaussian noise to the unlabeled data $\bm X^q_\text{UL}$, where noise $\sim\mathcal{N}(0, \sigma^2)$ and $\sigma = 0.2$. The perturbed data $\hat{\bm X}^q_\text{UL}$ is input to model $f^{\text{Swin}}{\phi}(\cdot)$ to obtain probability outputs $f^{\text{Swin}}{\phi}(\hat{\bm X}^q_\text{UL}) \in \mathbb{R}^{2 \times n_h \times n_w}$.

We also observe that the fetal head size in the first trimester is smaller compared to the second or third trimester, resulting in a large proportion of background pixels in US images. The consistency of these background pixels is unessential for model training. Therefore, we adopt a downsampling layer $\Downsampling(\cdot)$ to filter out insignificant pixels in $f^{\text{Swin}}_{\phi}(\hat{\bm X}^q_\text{UL})$ and $f^{\text{UNX}}_{\theta}(\bm X^q_\text{UL})$, prioritizing the representation of the fetal head in the image. The output size of $\Downsampling(\cdot)$ is denoted as $C_\text{ds}$, so the output is expressed as: $\Downsampling(f^{\text{UNX}}_{\theta}(\bm X^q_\text{UL})) \in \mathbb{R}^{2 \times C_\text{ds} \times C_\text{ds}}$ and $\Downsampling(f^{\text{Swin}}_{\phi}(\hat{\bm X}^q_\text{UL})) \in \mathbb{R}^{2 \times C_\text{ds} \times C_\text{ds}}$.
Furthermore, we normalize them in the channel dimension to induce sparsity in the features and enhance the model's resilience to perturbations. The consistency learning loss $\mathcal L_\text{con}$ is defined as:
\vspace*{-0.5em}
\[
\label{conloss}
\begin{split}
\mathcal{L}_{\text {con}} = \MSE( \mathcal{N}_{\text{L2}} (\Downsampling(f^{\text{UNX}}_{\theta}(\bm X_\text{UL}))) - \mathcal{N}_{\text{L2}}(\Downsampling(f^{\text{Swin}}_{\phi}(\hat{\bm X}^q_\text{UL})))))
\end{split}
\]
where $\MSE(\cdot)$ is mean squared error and $\mathcal{N}_{\text{L2}}(\cdot)$ introduces $\text{L}_2$ regularization.

\vspace*{-1em}
\subsection{The Overall Training Loss}
\label{overall_loss}

The overall training loss is a joint loss with three parts: the supervision loss $\mathcal L_\text{sup}$ for the labeled data, the semi-supervision loss $\mathcal L_\text{semi}$, and consistency loss $\mathcal L_\text{con}$ for the unlabeled data. All training losses are highlighted by a red dashed line in Fig.~\ref{fig:framework}. 
Mathematically, the joint loss can be expressed as:
\begin{equation}
\label{loss}
\mathcal{L}_{\text {total}}=(\mathcal{L}_{\text {sup }}^1+\mathcal{L}_{\text {sup }}^2)+\lambda(\mathcal{L}_{\text {semi }}^1+\mathcal{L}_{\text {semi }}^2)+\mathcal{L}_{\text {con}}
\end{equation}
where $\lambda$ represents the weighting factor for a Gaussian ramp-up function used exclusively for the unlabeled training set~\cite{Laine:2017}. This Gaussian ramp-up function facilitates the gradual transition of $f^{\text{UNX}}_{\theta}(\cdot)$ and $f^{\text{Swin}}_{\phi}(\cdot)$ from being initialized with the labeled training set to prioritizing learning from the unlabeled training set. The $\lambda$ is expressed as $\lambda = e^{-5 \times\left(1-t_{\text{Iter}} / t_{\text{Max Iter}}\right)^2}$, where $t_\text{Iter} = N_\text{epoch} / B$ and $B$ stands for the batch size.
$\mathcal{L}^1_\text{sup}$ and $\mathcal{L}^2_\text{sup}$ are the supervision losses for $f^{\text{UNX}}_{\theta}(\cdot)$ and $f^{\text{Swin}}_{\phi}(\cdot)$ based on the labeled data $\bm{X}_\text{L}$. $\mathcal{L}_\text{sup}$ is a combination loss, as follows:

\vspace*{-1.1em}
\begin{equation}
\begin{split}
\mathcal{L}_{\text {sup }}^1 &=\CE\left(f^{\text{UNX}}_{\theta}\left(\bm{X}_\text{L}\right), \bm{Y}_{\mathrm{gt}}\right)+\Dice\left(f^{\text{UNX}}_{\theta}\left(\bm{X}_\text{L}\right), \bm{Y}_{\mathrm{gt}}\right) \\
\mathcal{L}_{\text {sup }}^2&=\CE\left(f^{\text{Swin}}_{\phi}\left(\bm{X}_\text{L}\right), \bm{Y}_{\mathrm{gt}}\right)+\Dice\left(f^{\text{Swin}}_{\phi}\left(\bm{X}_\text{L}\right), \bm{Y}_{\mathrm{gt}}\right)
\end{split}
\end{equation}
\noindent $\CE(\cdot)$ and $\Dice(\cdot)$ denote the \ac{CE} loss and the standard \ac{Dice} loss, respectively. The semi-supervision losses are computed using pseudo labels $\tilde{Y}^\text{UNX}$ and $\tilde{Y}^\text{Swin}$ during co-training between $f^{\text{UNX}}_{\theta}(\cdot)$ and $f^{\text{Swin}}_{\phi}(\cdot)$, respectively. Therefore, the loss is defined as:

\vspace*{-1.2em}
\[
\label{semi_loss}
\begin{split}
\mathcal{L}_{\text {semi}}^1=\CE\left(f^{\text{UNX}}_{\theta}\left(\boldsymbol{X}_\text{UL}\right), \bm \tilde{Y}^\text{Swin}\right) + 
\Dice\left(f^{\text{UNX}}_{\theta}\left(\boldsymbol{X}_\text{UL}\right), \bm \tilde{Y}^\text{Swin}\right) \\
\mathcal{L}_{\text {semi}}^2=\CE\left(f^{\text{Swin}}_{\phi}\left(\boldsymbol{X}_\text{UL}\right), \bm \tilde{Y}^\text{UNX}\right) +
\Dice\left(f^{\text{Swin}}_{\phi}\left(\boldsymbol{X}_\text{UL}\right), \bm \tilde{Y}^\text{UNX}\right)
\end{split}
\]
The consistency learning loss $\mathcal{L}_{\text {con}}$ is explained in Section~\ref{ssl}.

\vspace*{-0.6em}
\section{Experiments and Results}
\vspace*{-0.6em}
\subsection{Datasets}

Two public fetal US image datasets for different scenarios are used to evaluate our proposed method. Both of them were acquired from healthy singleton pregnancies.
{\bf Spanish trans-cerebellum (ES-TCB)} dataset is acquired from two centers in Spain~\cite{Alzubaidi:2023} when pregnant women are in their $2^\text{nd}$ and $3^\text{rd}$ trimesters undergoing routine examinations~\cite{Alzubaidi:2023}. Students, experienced physicians, and radiologists annotated the region of interest (ROI) of the fetal head in all images. We use 681 annotated trans-cerebellum images to undertake our experiments.
{\bf HC18} images are captured by experienced sonographers utilizing two different US machines in the Netherlands~\cite{Heuvel:2018_b}. We use 999 annotated images in this study. The annotations represent the ROI that best fits the circumference of the fetal head in each image.
We use a 50:50 ratio for both datasets to split the data into training and test sets. In our experiments, we examine two scenarios with limited labeled data available: 5\% and 10\%. The number of images for training, validation, and testing is illustrated in Table \ref{data_profile}.
Data augmentation techniques are only applied to the labeled data during training. 
The detailed parameters of the data augmentation techniques are in our code repository on GitHub.

\begin{table}
\centering
\caption{The details of HC18 and ES-TCB datasets.}
\resizebox{\linewidth}{!}{%
\label{data_profile}
\setlength{\tabcolsep}{10pt}
\begin{tabular}{|lcccc|}
\hline
Name & Image Size & Train & Validation & Test \\
\hline
HC18 & $800\times540$ & 500 & 50 & 449 \\
ES-TCB & $959\times661$ & 342 & 50 & 289 \\
\hline
\end{tabular}%
}
\vspace*{-0.7em}
\end{table}

\vspace*{-0.7em}
\subsection{Benchmark Methods}

We compare the segmentation performance of SSL-EGAD with the benchmark and SOTA methods in SSL field.
There are nine SSL methods trained using the same hyperparameter settings. The test, labeled, and unlabeled training sets are randomly selected once and evaluated collectively with all methods. These SSL methods reported include: Mean Teachers (MT)~\cite{Tarvainen:2017}, Deep Co-Training (DCT)~\cite{Qiao:2018}, Uncertainty-Aware Mean Teachers (UAMT)~\cite{Yu:2019}, Cross Pseudo Supervision (CPS)~\cite{Chen:2021}, 
Cross Teaching between CNN and Transformer (CTCT)~\cite{Luo:2022}, Lightweight Model for Contrastive Learning (LMCT)~\cite{Wang:2025b}, and Pixel-level contrastive and cross-supervised (PCCS)~\cite{Ma:2024}.
For AL query strategy, we compare two of the most well-known and commonly used methods in the AL field, diversity-based (Random) and covering uncertainty-based (Entropy~\cite{Wang:2017}). 

\vspace*{-0.9em}
\subsection{Implementation Details and Evaluation Metrics}
We trained each method with 400 epochs. With the query strategy, each AL iteration was trained with 400 epochs, and the max iterative loop was set to 5 to ensure fair comparison under equal annotation budgets. The labeled batch size is set to 1, the unlabeled batch size is set to 4, the optimizer is stochastic gradient descent (SGD), the learning rate is set to 0.01, momentum is 0.9, and weight decay is 0.0001. For AL, we set iteration $i=5$. The size of $X^i_{\text{S}}$ is 1\% and 2\% of the labeled data in the 5\% and 10\% scenarios, respectively. Our code is developed in Python 3.11.5 using Pytorch 2.1.2 and CUDA 12.2 using one NVIDIA 4090 GPU. In training, we evaluate the UNeXt model on the validation set every epoch and store the weights of the best-performing UNeXt model. All images are resized to dimensions of $448 \times 448$ pixels. The above setting is also applied to all other benchmark methods directly without any modification.
For model evaluation, we utilize the following evaluation metrics: Dice Score (DSC), and difference measure: Hausdorff Distance (HD). We assess the performance of SSL-EGAD across scenarios involving 5\% and 10\% of labeled data for training.

\vspace*{-0.9em}
\subsection{Results}
\label{res}


The quantitative results of fetal head segmentation are summarized in Table~\ref{overall_res}. The first two rows show the fully supervised baselines, UNeXt and Swin-Unet. Under both 5\% and 10\% labeled data settings, our proposed SSL-EGAD consistently surpasses all other SSL methods and AL strategies, while maintaining results comparable to the fully supervised Swin-Unet. It effectively bridges the gap between UNeXt and Swin-Unet, highlighting the ability of our approach to generate a robust, lightweight model by efficiently exploiting unlabeled data. Overall, the proposed EGAD strategy proves more effective than conventional active learning strategies such as Random and Entropy, yielding superior segmentation accuracy and boundary precision.

\begin{table}
\centering
\caption{The quantitative results of SSL-EGAD (UNeXt) and other methods when 5\% and 10\% of label data for training. The best results are in {\bf bold}.
}
\resizebox{\linewidth}{!}{%
\label{overall_res}
\begin{tabular}{|l|c|c|cc|cc|}
\hline
\multirow{2}{*}{Method} & AL & \#  & \multicolumn{2}{c|}{HC18} & \multicolumn{2}{c|}{ES-TCB} \\
\cline{4-7}
& Strategy & Labeled & DSC $\uparrow$ & HD $\downarrow$ & DSC $\uparrow$ & HD $\downarrow$ \\
\hline
UNeXt & \multirow{2}{*}{N/A} & 100\% & 93.37 & 48.60 & 94.69 & 45.57 \\
Swin-Unet &  & 100\% & 96.94 & 12.51 & 96.96 & 14.91 \\
\hline
MT~\cite{Tarvainen:2017} & \multirow{8}{*}{N/A} & \multirow{8}{*}{5\%} & 89.43 &  59.26 & 90.81 & 58.83 \\
DCT~\cite{Qiao:2018} & & & 91.69 & 54.29 & 90.76 & 56.11 \\
UAMT~\cite{Yu:2019} & & & 90.97 &  57.57 & 89.98 & 59.39 \\
CPS~\cite{Chen:2021} & & & 91.35 & 58.51 & 91.42 &  51.15 \\
CTCT~\cite{Luo:2022} & & & 91.50 &  54.26 & 90.88  & 53.38 \\
PCCS~\cite{Ma:2024}  & & & 92.24 & 51.23 & 91.38 &  50.36 \\
LCMT~\cite{Wang:2025b}  & & & 93.24 & 46.90 & 92.70 &  47.38 \\
\hline
LCMT~\cite{Wang:2025b} & Random & \multirow{3}{*}{1\% $\times$ 5} & 94.06 & 35.87 & 91.84 & 57.59 \\
CEAL~\cite{Wang:2017} & Entropy &  & 94.07 & 41.13 & \textbf{94.50}  & 48.82 \\
SSL-EGAD & EGAD &  & \textbf{95.09} & \textbf{28.77} & 94.04  & \textbf{41.43} \\
\hline
MT~\cite{Tarvainen:2017} & \multirow{8}{*}{N/A} & \multirow{8}{*}{10\%} & 94.38 &  35.19 & 93.20 &  48.22 \\
DCT~\cite{Qiao:2018} & & & 94.67 & 35.25 & 93.93 & 45.07 \\
UAMT~\cite{Yu:2019} & & & 94.86 &  32.93 & 93.07 & 50.90 \\
CPS~\cite{Chen:2021} & & & 94.61 & 33.60 & 92.79 & 50.15 \\		
CTCT~\cite{Luo:2022} & & & 94.80 &  35.80 & 94.38 &  43.69 \\
PCCS~\cite{Ma:2024}  & & & 95.13 & 29.80 & 94.83 &  44.11 \\
LCMT~\cite{Wang:2025b} & & & 95.31 & 29.50 & 95.15 & 39.32 \\
\hline
LCMT~\cite{Wang:2025b} & Random & \multirow{3}{*}{2\% $\times$ 5} & 95.67 & 28.68 & 95.35 & 36.39 \\
CEAL~\cite{Wang:2017} & Entropy &  & 95.69 & 29.17 & 95.40 & 42.24 \\
SSL-EGAD & EGAD &  & \textbf{96.44} & \textbf{19.15} & \textbf{96.21}  & \textbf{25.74} \\
\hline
\end{tabular}
}
\vspace*{-0.7em}
\end{table}


\textbf{Ablation Study.} We conducted an ablation study to analyze the contribution of each component in our framework using 5\% labeled data from the HC18 dataset. As shown in Table~\ref{ablation_al}, the SSL baseline surpasses the fully supervised model, indicating the benefit of leveraging unlabeled data. Comparing the second and third rows reveals that combining CNN and Transformer backbones leads to better feature representation than using dual CNN architectures. Incorporating consistency learning ($\mathcal{L}_{\text {con}}$) further enhances segmentation performance, demonstrating its effectiveness in stabilizing training and improving generalization. Finally, integrating our proposed EGAD strategy yields the best results across both datasets, confirming its effectiveness in selecting informative samples and fully exploiting unlabeled data.

\begin{table}
\centering
\caption{Ablation study on the effectiveness of each component using 5\% labeled data. The best results are in {\bf bold}.
}\label{ablation_al}
\resizebox{\linewidth}{!}{%
\begin{tabular}{|l|c|c|c|}
\hline
\multirow{2}{*}{Method} & \multirow{2}{*}{Backbone} & \multicolumn{2}{c|}{DSC $\uparrow$}\\
\cline{3-4}
& & {HC18} & {ES-CB}\\
\hline
Fully Supervised & UNeXt & 38.79 & 81.24 \\
SSL & Swin-Unet+UNeXt & 88.85 & 88.40 \\
SSL+$\mathcal{L}_\text{con}$ & UNeXt+UNeXt & 85.88 & 87.73 \\
SSL+$\mathcal{L}_\text{con}$ & Swin-Unet+UNeXt & 93.22 & 92.95 \\
SSL+$\mathcal{L}_\text{con}$+EGAD & Swin-Unet+UNeXt & \textbf{95.09} & \textbf{94.04} \\
\hline
\end{tabular}
}
\vspace*{-1em}
\end{table}

\vspace*{-0.5em}
\section{Conclusion}


In this study, we propose SSL-EGAD, a framework integrating AL with SSL to overcome limitations of existing methods in fetal head segmentation. By guiding a lightweight model to focus on informative regions, SSL-EGAD improves structural understanding and generalization. Experiments show that SSL-EGAD achieves superior performance with minimal labeled data, reducing annotation costs while maintaining high accuracy in fetal head analysis. The EGAD strategy is task-agnostic and can be integrated with other SSL pipelines. Extending the framework to additional fetal structures and imaging modalities will be explored in future work.

\newpage

\section{COMPLIANCE WITH ETHICAL STANDARDS}

This research study was conducted retrospectively using human subject data made available in open access by \cite{Heuvel:2018_b,Alzubaidi:2023}. Ethical approval was not required as confirmed by the license attached with the open access data.
\vspace*{-0.75em}
\bibliographystyle{IEEEbib}
\bibliography{strings,refs}

\end{document}